# Inverse Resource Rational Based Stochastic Driver Behavior Model

Mehmet F. Ozkan*,a, Yao Ma**,a

*mehmet.ozkan@ttu.edu, **yao.ma@ttu.edu
a Department of Mechanical Engineering, Texas Tech University, Lubbock, TX 79409 USA

Abstract: Human drivers have limited and time-varying cognitive resources when making decisions in real-world traffic scenarios, which often leads to unique and stochastic behaviors that can not be explained by perfect rationality assumption, a widely accepted premise in modeling driving behaviors that presume drivers rationally make decisions to maximize their own rewards under all circumstances. To explicitly address this disadvantage, this study presents a novel driver behavior model that aims to capture the resource rationality and stochasticity of the human driver's behaviors in realistic longitudinal driving scenarios. The resource rationality principle can provide a theoretic framework to better understand the human cognition processes by modeling human's internal cognitive mechanisms as utility maximization subject to cognitive resource limitations, which can be represented as finite and time-varying preview horizons in the context of driving. An inverse resource rational-based stochastic inverse reinforcement learning approach (IRR-SIRL) is proposed to learn a distribution of the planning horizon and cost function of the human driver with a given series of human demonstrations. A nonlinear model predictive control (NMPC) with a time-varying horizon approach is used to generate driver-specific trajectories by using the learned distributions of the planning horizon and the cost function of the driver. The simulation experiments are carried out using human demonstrations gathered from the driver-in-the-loop driving simulator. The results reveal that the proposed inverse resource rational-based stochastic driver model can address the resource rationality and stochasticity of human driving behaviors in a variety of realistic longitudinal driving scenarios.

## 1. INTRODUCTION

Automated vehicles (AVs) are widely expected to deploy in human-dominated traffic scenarios where human-operated vehicles and AVs share the road with frequent interactions in the foreseeable future (Di and Shi, 2021). The driver behavior model is necessary for AVs to understand the intentions of drivers to safely interact with human-operated vehicles in highly interactive human-dominated traffic scenarios. However, existing driver behavior models, such as inverse reinforcement learning (IRL) based models, assume that human drivers are perfect rational decision-makers during the operation of vehicles and consider that human drivers make actions with a fixed planning horizon setting in traffic (Kuderer et al., 2015) (Sadigh et al., 2016) (Gao et al., 2018) (Naumann et al., 2020) (Ozkan and Ma, 2022). However, this assumption is not realistic because of human drivers' behavioral stochasticity and limited cognitive resources to make a decision in a time-varying manner. To this end, it is crucial to develop a comprehensive driver behavior model that addresses the human drivers' cognitive processes with the rational use of limited resources and stochastic driving behaviors in real-world driving scenarios.

Resource rationality, a principle in cognitive science for modeling the human decision-making mechanism as utility maximization while taking into account cognitive constraints, can offer a theoretical framework to better understand human behavior by incorporating realistic assumptions about human agents' cognitive resource limitations (Griffiths et al., 2015) (Lieder and Griffiths, 2020). Resource rationality provides several opportunities to more precisely understand human behavior, such as making generalizable predictions about how long it will take humans to make decisions and how much information humans will include when making decisions (Ho and Griffiths, 2022). A recent study has been proposed to understand the human inverse model of resource rational processes with the IRL framework (Zhi-Xuan et al., 2020). The proposed method considers that the human agents have limited and time-varying resources when making decisions and aims to learn the goals and preferences of the human agents with the learned reward functions while considering the resource rationality.

Motivated by the discussions above, this work aims to develop an inverse resource rational-based stochastic driver behavior model that reflects the resource rationality and stochasticity of human behavior in real-world longitudinal driving scenarios. This study makes the following contributions: 1) an inverse resource rational-based stochastic inverse reinforcement learning approach is developed to learn the dynamic planning horizon and cost function distributions of the human driver from human driving demonstrations. 2) the developed inverse resource rational-based stochastic driver behavior model is employed to compute driver-specific trajectories by using the nonlinear model predictive control with the time-varying horizon approach in realistic longitudinal driving scenarios. To the best of the authors' knowledge, this is the first study explicitly addressing the resource rationality of human drivers' cognitive processes in real-world driving scenarios.

The remainder of this work is structured as follows. Section 2 develops the inverse resource rational-based stochastic driver behavior model. The numerical findings of the developed driver behavior model are presented in Section 3. The paper is concluded in Section 4.

## 2. DRIVER BEHAVIOR MODEL

This study aims to acquire the planning horizon and cost function distributions that best characterize the human driver's cognitive process while operating vehicles in traffic. To this end, we will first outline a stochastic inverse reinforcement learning approach (SIRL) approach to learn the cost function distribution of the human driver with a fixed planning horizon, and we will then integrate the inverse resource rationality to acquire the planning horizon distribution of the human driver with a given set of driving demonstrations. At last, we will introduce the nonlinear model predictive control (NMPC) algorithm with a time-varying horizon to generate driver-specific vehicle trajectories in longitudinal driving scenarios by using the learned planning horizon and cost function distributions. Fig. 1 shows the schematic of the developed driver behavior model framework.

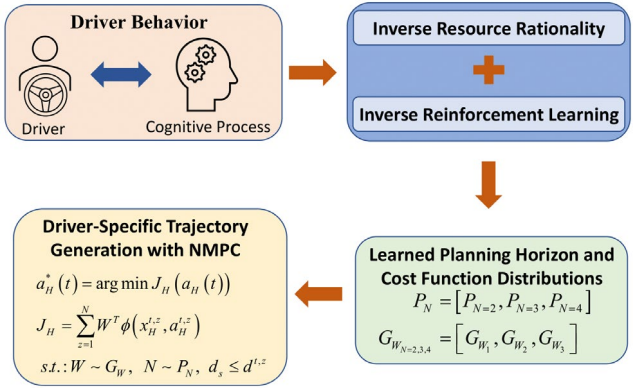

Fig. 1. Schematic of the developed driver behavior learning framework.

### 2.1 Cost Function Distribution Learning with Stochastic Inverse Reinforcement Learning (SIRL)

The primary goal of the SIRL approach is to learn a driver's cost function distribution from a set of human demonstrations $D$ consisting of $K$ observed trajectory segments in which each cost function from the distribution reflects the driver's demonstrated behavior for each seen trajectory segment. The cost function can be specified as the weighted features:

$$J_i = W_i^T \phi_i(r_i) \quad (1)$$

where $J_i$ is the cost function, $W_i$ is the weight vector, and $\phi(r_i) = (\phi_1, \phi_2, \cdots, \phi_n)^T$ is the feature vector that consists of $n$ features for each observed trajectory segment; $r_i$ defines the longitudinal position representation of each trajectory segment, and subscript $i$ represents the $i$th trajectory segment in the human demonstrations.

We employ quintic polynomials as a finite-dimensional representation of longitudinal vehicle trajectories. Because of the advantages of smooth motion, quick calibration, and light computing, quintic polynomials have been frequently employed for vehicle motion planning problems in existing research such as autonomous vehicle trajectory planning and driver behavior learning (Kuderer et al., 2015) (Ozkan et al., 2021). Hence, the longitudinal vehicle trajectory model is specified as a one-dimensional quantic polynomial, and the longitudinal position representation of each trajectory segment $i$ is stated as:

$$r_i(t) = y_0 t^5 + y_1 t^4 + y_2 t^3 + y_3 t^2 + y_4 t + y_5 \quad (2)$$

where $y_{0-5}$ are the coefficients for each demonstrated trajectory segment; $t \in [t_i, t_i + H]$ and $H$ is the length of each trajectory segment. The longitudinal velocity and acceleration can be expressed by using the derivatives of $r_i(t)$ such as $\dot{r}_i(t)$ and $\ddot{r}_i(t)$, respectively.

The goal is to find the optimal weight vector $W^*$ that maximizes the likelihood of the demonstrations for each observed trajectory segment:

$$W_i^* = \arg\max_{W_i} p(D_i | W_i) = \arg\max_{W_i} \prod_{k=1}^{B} p(r_i | W_i) \quad (3)$$

where $B$ defines the number of the fixed planning horizons within each observed trajectory segment and each trajectory subsegment has the same planning horizon $N$, $p(r|W)$ specifies a probability distribution across the trajectory segment that is proportional to the negative exponential costs determined using the Maximum Entropy principle along the trajectory segment. (Ziebart et al., 2008):

$$p(r_i | W_i) = \exp(-W_i^T \phi(r_i)) \quad (4)$$

The weight vector $W$ can be derived with the gradient of the optimization problem. The gradient can be acquired by subtracting the observed feature values from the expected feature values:

$$\nabla \phi_i = \tilde{\phi}_i - \phi_i^e \quad (5)$$

The feature values of the most likely trajectory can be used to determine the expected feature values:

$$\phi^e \approx \phi_i \left( \arg\max_{r_i} p(r_i | W_i) \right) \quad (6)$$

The gradient descent approach can be used to update the feature weight vector with the learning rate $\alpha$ for each trajectory segment:

$$W_i \leftarrow W_i - \alpha \nabla \phi_i \quad (7)$$

After the derivation of the weight vector $W$, the next step will be generating a distribution from the learned set of $K$ different cost functions. We used t Copula approach (Bouye et al., 2000) to fit the learned set of feature vectors $W = [W_1, W_2, W_3, ..., W_K]$ in a multivariate distribution $G_W$. For more details about the cost function distribution fitting process, the reader is referred to (Ozkan et al., 2021).

2.2 Feature Design

The features listed below are utilized to represent the key properties of longitudinal driving behaviors:

**Acceleration:** Capturing riding comfort along the longitudinal direction:

$$\phi_a(t) = \int_t^{t+N} \|\ddot{r}(t)\|^2 \, dt \qquad (8)$$

**Desired Speed:** Reaching and maintaining the traffic speed limit $v_d$:

$$\phi_{ds}(t) = \int_t^{t+N} \|v_d - \dot{r}(t)\|^2 \, dt \qquad (9)$$

**Relative Speed:** Maintaining a constant gap distance while observing the speed of the preceding vehicle $v_p$:

$$\phi_{rs}(t) = \int_t^{t+N} \|v_p(t) - \dot{r}(t)\|^2 \, dt \qquad (10)$$

**Steady Car-following Gap Distance:** Achieving a desired car-following gap distance $d_c$ with the constant time headway approach:

$$d_c = \dot{r}(t)\tau + d_s \qquad (11)$$

$$\phi_{cd}(t) = \int_t^{t+N} \|d(t) - d_c\|^2 \, dt \qquad (12)$$

where $d(t)$ is the car following gap distance to the preceding vehicle at the time $t$, $\tau$ is the observed minimum time headway of the human driver with the given human demonstrations and $d_s$ is the standstill distance.

**Safe Interaction Gap Distance:** Maintaining a safe car-following gap distance when closely following the preceding vehicle in congested traffic:

$$\phi_{sd}(t) = \int_t^{t+N} \|d(t) - d_s\|^2 \, dt \qquad (13)$$

**Free Motion Gap Distance:** Capturing the driver's desired gap distance to the preceding traffic when the driver controls the vehicle in free motion:

$$\phi_{fd}(t) = \int_t^{t+N} e^{-d(t)} dt \qquad (14)$$

2.3 Driving Conditions and Feature Selection

We classified the observed trajectory segments depending on their observed driving conditions, and the distinct sets of features are applied to each driving condition. Therefore, we considered three longitudinal different driving conditions, and the corresponding features are applied to each driving condition. These three driving conditions are briefly outlined below.

**Steady car-following:** The steady car-following driving phase is a driving condition in which the average time headway $(\text{THW} < 6 \text{ s})$ (Vogel, 2003) and average time to collision inverse $(\text{TTCi} < 0.05 \text{ s}^{-1})$ (Lu et al., 2010) for each trajectory segment. The features $\phi_a$, $\phi_{ds}$, $\phi_{rs}$ and $\phi_{cd}$ are employed to describe the driver's steady car-following condition.

---

**Algorithm 1: Inverse resource rational-based stochastic driver behavior learning algorithm**

**Input:** $(r_1, r_2, ... r_K)$, $N_{total} = [2\text{ s}, 3\text{ s}, 4\text{ s}]$

**Output:**
$G_{W_{N=2,3,4}} = [G_{W_1}, G_{W_2}, G_{W_3}]$, $P_N = [P_{N=2}, P_{N=3}, P_{N=4}]$,
$(r_{1,1}^*, r_{1,2}^*, ..., r_{1,B}^*, r_{2,1}^*, r_{2,2}^*, ..., r_{2,B}^*, ..., r_{K,1}^*, r_{K,2}^*, ..., r_{K,B}^*)$

1:   Classify the trajectory segments depending on their driving conditions
2:   **for** each classified trajectory segment **do**
3:     Initialize weight pool $W_{j=1,2,3} \leftarrow [\ ]$
4:     **for** all trajectory segments **do**
5:      **for** each planning horizon in $N_{total}$ **do**
6:       Partition each trajectory segment into $B$ planning subsegments $(r_{j,1}, r_{j,2}, ..., r_{j,B})$
7:       $W_i \leftarrow$ all-ones vector
8:       $\tilde{\phi}_i = \frac{1}{B}\sum_{k=1}^{B}\phi_i(\tilde{r}_{i,k})$
9:       **while** $W_i$ not converged **do**
10:        **for** all $r_{i,k} \in (r_{i,1}, r_{i,2}, ..., r_{i,B})$ **do**
11:         $(y_5, y_4, y_3) \leftarrow$ (position, velocity, acceleration)
12:         at the initial state of the $r_{i,k}$
13:         Optimize $(y_2, y_1, y_0)$ with respect to $W_i^T \phi_i$
14:        **end for**
15:        Update $W_i$ with respect to the gradient of the optimization problem $\nabla \phi_i$
16:       **end while**
17:       $W \leftarrow W_i$
18:       Record the final gradient $\nabla \phi_i$
19:      **end for**
20:     Record the planning horizon value $N_{total_{\min \nabla \phi_i}}$ and weight vector $W_{i_{\min \nabla \phi_i}}$ with the lowest final gradient
21:     **end for**
22:   **end for**
23:   $G_{W_j} \leftarrow$ Fit each cluster's set of weight vectors $W_{j=1,2,3}$ with the corresponding planning horizon $N_{total_{\min \nabla \phi_i}}$ into t Copula distribution
24:   $P_N \leftarrow$ Calculate the probability vector of the planning horizons based on their occurrences

---

**Free motion:** The driver operates the vehicle without engaging with the preceding vehicle during the free motion driving condition. Based on the numerical analysis by (Ozkan et al., 2021), the following conditions are defined to represent the free motion driving condition. The features $\phi_a$, $\phi_{ds}$ and $\phi_{fd}$ are utilized for the free motion driving condition.

1. The average THW > 6 s and average TTCi ≤ 0 s$^{-1}$.
2. The average car-following gap distance > 35 m
3. The average driver's speed > 5 m/s

**Unsteady car-following:** The driver operates the vehicle in the unsteady car-following condition when the human driver is neither in steady car-following nor in free motion driving conditions. The features $\phi_a$, $\phi_{ds}$, $\phi_{rs}$ and $\phi_{sd}$ are utilized to identify the driver's unsteady car-following condition.

To ensure that all features are equally sensitive, we normalized each feature to the range of $[0,1]$.

2.4 Algorithm Implementation of Inverse Resource Rational Based Stochastic Driver Behavior Model

In the previous sections, we outlined the main components of the SIRL approach for learning the cost function distribution of the human driver with the fixed planning horizon setting. Now, we will integrate the SIRL with the inverse resource rationality to learn the planning horizon distribution of the human driver. In the inverse resource rationality approach, the goal is to find the distribution of the optimal planning horizon values where each optimal horizon value gives the most likely trajectory against the ground truth during the learning process. By this, the set of the planning horizon values during the learning process is defined as $N_{total} = [2\,s, 3\,s, 4\,s]$. The details of the driver behavior learning process are included in Algorithm 1.

2.5 Trajectory Generation with Nonlinear Model Predictive Control (NMPC)

In the previous section, we learned the planning horizon and cost function distributions that best represent the driver's cognitive process when operating vehicles in traffic by using human demonstrations. We will then use the learned planning horizon and cost function distributions to compute driver-specific motions in longitudinal driving scenarios with the NMPC algorithm with a time-varying horizon approach. In the longitudinal driving scenario, we consider that the human driver can adequately estimate the movements of the preceding traffic if the preview time horizon is relatively short (Sadigh et al., 2016) (Ozkan and Ma, 2022). At each time step $t$, the optimization problem needs to be solved sequentially over the prediction time horizon $N$ to compute the driver-specific motions:

$$a_H^*(t) = \arg\min J_H(a_H(t))$$
$$J_H = \sum_{z=1}^{N} W^T \phi(x_H^{t,z}, a_H^{t,z}) \quad (15)$$
$$s.t.: W \sim G_{W_{N=2,3,4}},\ N \sim P_N,\ d_s \leq d^{t,z},\ v_{min} \leq v^{t,z} \leq v_{max}$$

where $a_H^*(t)$ is the optimal acceleration; $x_H^{t,z}$ and $a_H^{t,z}$ are the $(t+z)^{th}$ predicted vehicle state and acceleration, respectively; $\phi$ is the feature vector depending on the observed driving conditions over the planning horizon; $W$ is the random feature weight vector sample from the distribution $G_W$; $N$ is the random prediction horizon sample from the learned planning horizon distribution $P_N$ and $v_{min}$ and $v_{max}$ are the minimum and maximum speed constraints.

3. RESULTS AND DISCUSSION

3.1 Driver Behavior Model Development

We used our previously developed driver-in-the-loop driving simulator (Ozkan et al., 2021) to collect realistic driving data for developing the proposed driver behavior model. The simulation scenarios concentrate on car-following scenarios, in which a driver follows a preceding vehicle on the road that operates at various specified realistic speed trajectories which include cruising, frequent stop-and-go, and transient traffic conditions. The driving data set is collected using nine different driving scenarios, each of which is repeated 30 times by the same driver. At a rate of 10 Hz, data is extracted from the simulation environment. The driver model is built using 270 leader-follower trajectories.

To evaluate the effectiveness of the proposed driver behavior model in different longitudinal driving settings, 25 trajectories for each driving scenario are randomly chosen for training, and the remaining five trajectories for each driving scenario are applied for testing. For the trajectory optimization process for the quintic polynomial parameters during the learning, as mentioned in step 13 of Algorithm 1, the BFGS Quasi-Newton method (Fletcher, 1987) is used. The length of each trajectory segment $H$ is set to 12 s. The standstill distance $d_s$ and learning rate $\alpha$ are set to 5 m and 0.1, respectively, and $v_{min}$ is set to 0 m/s and $v_{max}$ is set to the maximum speed of the preceding vehicle during the trip. For the trajectory generation, the NMPC design in (15) is employed to generate 50 samples for each driving scenario.

3.2 Driver Behavior Model Evaluation

We will assess the proposed driver behavior model's performance under several driving scenarios in this section. First, we will assess the developed driver behavior model's training performance. Fig. 2 shows the $L^2$ norm of weight update gradients with different planning horizon settings for a trajectory segment that is used in the training process. During the optimization, it is shown that the feature weights converge after around 200 iterations for all planning horizon settings in this particular trajectory segment. Fig. 3 depicts speed trajectories of the trajectory segment with different planning horizons in training, including the initial guess, predicted trajectory, and ground truth. It can be seen that the predicted trajectories with all the planning horizon settings can approach ground truth when the weights converge to optima via incremental gradient updates.

Fig. 4 shows the learned probability distribution bar graph of the planning horizon values for three different driving scenarios from the demonstrations, along with preceding traffic speed trajectories of the corresponding driving scenarios. Notably, we can see that the proposed driver behavior model can learn different optimal planning horizon values for each driving scenario and the probability of each planning horizon value varies among different driving scenarios. These findings demonstrate that drivers have time-varying cognitive resources when making decisions in real-world traffic scenarios, and the proposed driver behavior model can capture the resource rationality of the human driver with the dynamic planning horizon approach.

We then assess the testing results of the developed IRR-SIRL driver behavior model. Fig. 5 and Fig. 6 illustrate the observed and predicted trajectory samples in one of the driving scenarios from the demonstrated driving data. The results reveal that the developed driver behavior model can generate various trajectory samples that reflect the complexity of the driver's individual driving strategies. The proposed IRR-SIRL model predicts accurate trajectories when compared to actual trajectory samples, demonstrating a high degree of validity.

Table 1 shows the RMSE values between the observed and the predicted trajectories for testing using the learned cost function. We observe that the proposed driver behavior model can mimic the demonstrated trajectories of the human driver with minor prediction errors. Since human drivers demonstrate highly uncertain driving behaviors due to their complex cognitive processes and the expected features are derived as the feature values of the most likely trajectories throughout the learning process, minor deviations between the observed and predicted trajectories are expected in testing.

The results collectively demonstrate that the proposed inverse resource rational-based stochastic driver behavior model can learn and mimic a driver's observed resource rational, distinct, and rich driving strategies in diverse longitudinal driving scenarios. The developed driver behavior model provides the possibility for automated vehicles to make better predictions about human drivers' behaviors while taking into account the realistic assumptions about human drivers' cognitive limitations in real-world traffic scenarios.

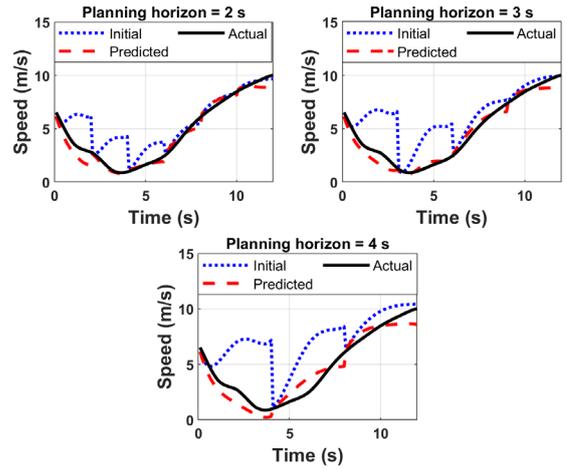

Fig. 3. Speed trajectories of a trajectory segment with different planning horizon values for training.

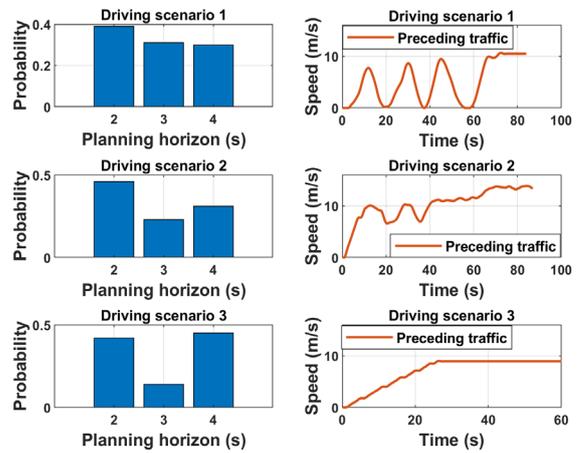

Fig. 4. Probability distribution bar graph of the planning horizon settings and preceding traffic trajectories of three different driving scenarios.

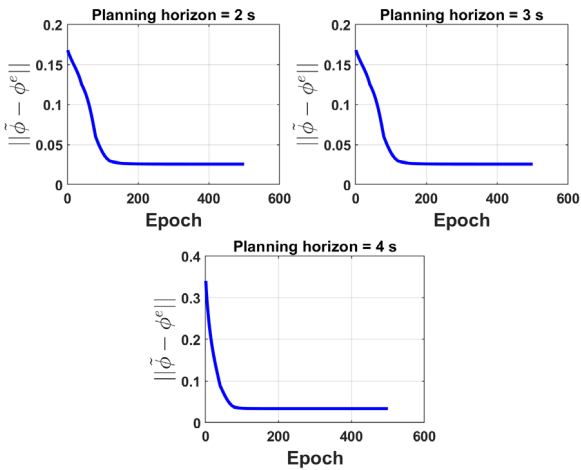

Fig. 2. Gradients ($L^2$ norm) of a trajectory segment with different planning horizon settings for training.

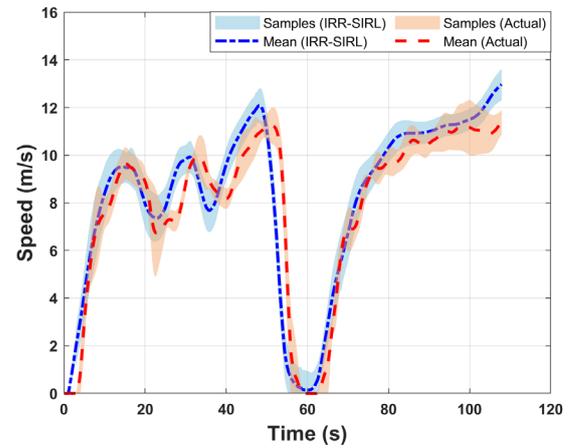

Fig. 5. Speed trajectories of a driving scenario in testing.

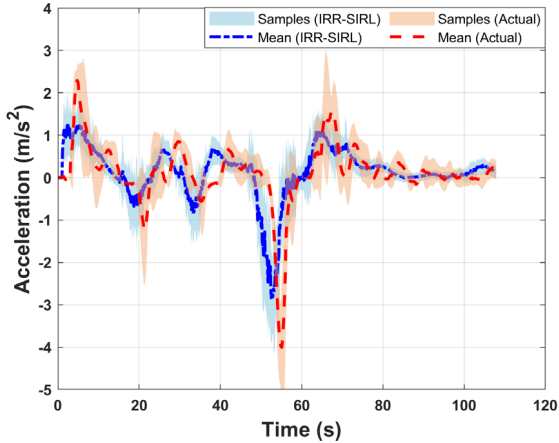

Fig. 6. Acceleration trajectories of a driving scenario in testing.

Table 1. Average RMSE values of IRR-SIRL driver behavior model in testing.

| Model | Speed (m/s) | Acceleration (m/s$^2$) |
|---|---|---|
| IRR-SIRL | 1.28 | 0.51 |

## 4. CONCLUSION AND FUTURE WORK

In this work, an inverse resource rational-based stochastic driver behavior model is developed to learn the human driver's resource rational and stochastic driving behaviors in longitudinal driving scenarios. The proposed driver behavior model employs an inverse resource rationality approach integrated with the inverse reinforcement learning framework to generate the planning horizon and cost function distributions of the human driver that captures the resource rationality and stochasticity of the human driver behavior with given human driving demonstrations. The numerical results indicate that the proposed driver behavior model can learn and mimic the human driver's resource rational, distinct, and rich driving behaviors across a wide range of longitudinal traffic scenarios.

The proposed driver behavior model is the first step towards understanding the resource utilization in the cognitive processes of human drivers when driving in traffic. We have so far assumed the possible planning horizon values can be enumerated and the optimization problem can be solved for each planning horizon setting in the learning process of the driver behavior model. We will further extend the proposed driver behavior model by including the planning horizon value as a decision variable in the inner optimization problem to get optimal planning horizon values during the learning process.